# First experimental demonstration of a scalable linear majority gate based on spin waves

Florin Ciubotaru[1], Giacomo Talmelli[1,2], Thibaut Devolder[3], Odysseas Zografos[1], Marc Heyns[1,2],
Christoph Adelmann[1], and Iuliana P. Radu[1]
[1] imec, Leuven, Belgium, email: Florin.Ciubotaru@imec.be
[2] KU Leuven, Leuven, Belgium, [3] Centre de Nanosciences et de Nanotechnologies, Univ. Paris-Sud, Orsay, France

*Abstract*—We report on the first experimental demonstration of majority logic operation using spin waves in a scaled device with an in-line input and output layout. The device operation is based on the interference of spin waves generated and detected by inductive antennas in an all-electrical microwave circuit. We demonstrate the full truth table of a majority logic function with the ability to distinguish between strong and weak majority, as well as an inverted majority function by adjusting the operation frequency. Circuit performance projections predict low energy consumption of spin wave based compared to CMOS for large arithmetic circuits.

## I. INTRODUCTION

Spintronic devices based on spin waves are promising alternatives to CMOS technology with high potential for power and area reduction per computing throughput [1,2]. The information can be encoded in either the amplitude or the phase of spin waves, while logic operation is based on their interference. Different spin-wave-based logic systems have been proposed, *e.g.* Mach Zender interferometers [3], magnonic transistors [4], or spin wave majority gates (SWMGs) [1]. The SWMG is the most promising concept as it possesses a higher expressive power than *e.g.* NAND or NOR gates and thus may reduce circuit complexity. The basic functionality of a such device has been proven at a mm scale using YIG films[5]. Device scalability to µm and nm dimensions has been predicted by micromagnetic simulations [6,7]. So far, the SWMG devices have been based on a "trident" shape. However, such a shape is difficult to scale due to increasing spin wave reflection at bends with µm and nm dimensions [6]. Moreover, such a shape is challenging to print at the nm scale using conventional lithography. In this work, we demonstrate a majority gate using spin wave interference in µm-sized ferromagnetic waveguides using a sequential "in-line" layout of input and output inductive antennas.

## II. FABRICATION AND RF PROPERTIES

The device consisted of a magnetic stripe that served as a waveguide for the spin waves. Three inductive antennas were used to excite spin waves and an addition antenna was used to detect the resulting wave after interference, *i.e.* after computation (see Fig. 1).

For the waveguide, a Ta(3nm)/CoFeB(30nm)/Ta(3nm) film stack was sputtered onto 300 nm of $SiO_2$ on Si (100). The stack was then patterned into stripes with a width of 4 µm using ion-beam etching. The ferromagnetic CoFeB acted as the waveguide for the spin waves while the Ta layers served as seed and cap layers to prevent from oxidation of the magnetic film. The waveguide was finally covered by 40 nm of $SiN_x$ for electric isolation (Fig. 2(a)). Subsequently, Ti/Au inductive antennas with a width of 500 nm were fabricated by electron-beam lithography and lift-off (Fig. 2(b)). The antennas were connected to microwave coplanar waveguides and contacted by RF picoprobes (Fig. 2 (c)). A vector network analyzer (VNA) was used for both the excitation and characterization of the spin waves.

Spin waves were generated in the waveguide by the Oersted field generated by the RF currents flowing in the U-shaped input antennas and detected by a single-wire output antenna. This design provided a weak electromagnetic parasitic coupling between adjacent input antennas (see Figs. 3(a) and (b)), as well as between the input and output antennas, as indicated both by experiments and electromagnetic simulations [8]. Electromagnetic simulations of the Oersted field distribution (Fig. 3(c)) show that the U-shaped antennas can efficiently excite spin waves with wavelengths down to 700 nm (Fig. 3(d)).

## III. SPIN WAVE PROPERTIES AND INTERFERENCE

Spin wave transmission experiments were performed with the CoFeB waveguide magnetized by a magnetic field transverse to its long axis. In a first step, the propagation characteristics of spin waves emitted from each of the three input antennas towards the output antenna were determined. A schematic of the experimental configuration is shown in Fig. 4(a); Fig. 4(b) shows a typical transmitted signal from input $I_1$ towards output O, corresponding to a spin wave propagation distance of 4.8 µm. The minimum spin wave frequency is given by the ferromagnetic resonance, whereas the maximum frequency depends on maximum wavenumber that an antenna can excite. The full frequency-field dependence of the transmitted signal from input $I_1$ to output O is shown in Fig. 4(c). The device allowed for the generation and propagation of spin waves in a wide frequency range between 3 GHz and 22 GHz, depending on the external applied field. The dispersion relation calculated using parameters extracted from experiments (see Fig.4 (d)) demonstrates that the minimum spin wave frequency matches well the ferromagnetic resonance frequency, while the upper limit was set by the maximum wavevector ($k_{max}$ ~ 8.9 rad/µm) that can be excited by the antenna (corresponding to a wavelength of $\lambda = 700$ nm). The spin wave transmission from inputs $I_2$ and $I_3$ to the output O is shown in Fig. 5. The dephasing due to the different propagation distances could clearly be observed.

Subsequent experiments studied the interference of the spin waves generated simultaneously by multiple input antennas (see Fig. 6(a)). Microwave currents with the same frequency were applied to all three antenna inputs. The output signal was studied as a function of the input frequency, bias field, and relative phase difference between the input signals. For a given set of field-frequency parameters, an oscillatory signal was detected by varying the phase of the input signals corresponding to the constructive or destructive interference of the three generated spin waves. For example, Fig. 6(b) shows the detected signal for a phase rotation of up to $4\pi$ of the signal at input $I_1$, while $I_2$ and $I_3$ were kept in phase. The position of the maxima/minima could be tuned by varying the applied frequency, which changes the spin wave wavelength and thus modifies the interference pattern.

## IV. LOGIC FUNCTIONS

Building logic functions based on spin wave interference requires the control of both the amplitude and the phase of the spin waves generated by each input. Signal matching at the output was obtained by phase shifters and attenuators in the microwave circuits of each input antenna (see Fig. 7(a)). Figure 7(b) shows that the amplitude and phase of the input signals could be synchronized over a 1 GHz bandwidth for a bias field of 40 mT.

Input 0 and 1 logic states of a SWMG were defined as the phase of the spin wave signals, *i.e.* as phases of 0 and $\pi$, respectively. The variation of the phase-sensitive S-parameter (here the imaginary part) measured by the VNA was used to define the logic output signal. Positive and negative variations correspond to output wave phases of $\pi$ and 0, respectively (logic 1 and 0, respectively). Using the phase of one input as a control signal, for example setting the phase of $I_3$ to $\pi$ (Fig. 8(a)), and changing the phase of $I_1$ and $I_2$ between 0 and $\pi$, a logic OR function could be demonstrated, as shown in Fig. 8(b). In addition, a logic AND gate could be demonstrated by setting the phase of the control input $I_3$ to 0.

By individually controlling the phase of each input, the truth table of the logic majority operation was demonstrated over a frequency bandwidth of ~300MHz, as shown in Fig. 9. Weak and strong majority states could be distinguished. The clear separation between the states suggests that adding additional inputs will allow to create an *n*-state logic. By tuning the applied frequency on the inputs, an oscillatory output signal was observed (Fig. 10(a)). This fact is explained by the variation of the global phase due to the dependence of spin wave wavelength on frequency, leading to a change of the interference pattern at the output for different frequencies. Thus, an inverted majority gate can be obtained in the same device by tuning the operation frequency (Fig. 10(b)).

### *Benchmarking*

Spin wave logic concepts, and more specifically spin wave devices (SWDs) [1] have been benchmarked several times [9, 10]. All results show that using efficient voltage-driven spin wave generation and detection, SWDs can outperform state-of-the-art CMOS technology in terms of energy consumption. To showcase this, based on previous benchmarking work [11], we adapt the energy calculations of the 10 designs described in Fig. 11. These benchmarks are combinational and represent a common subset of arithmetic designs used in digital integrated circuits. The energy consumption per operation of SWDs is compared in Fig. 12 to the 10nm CMOS technology node [12]. The energy consumption per operation of the SWD circuits is on average 7.6x times lower than for 10 nm CMOS. This benchmarking highlights the potential for ultralow-power logic built based on SWDs.

### *Weak & strong majority use cases*

As shown in Fig. 9, the device allows for the distinction of strong and weak majority signals. This can be efficiently exploited by non-boolean or multilevel computational techniques. More specifically, it has been shown that strong/weak majority distinction can be applied to signal processing, such as pattern recognition [13]. Moreover, the above capability can be useful in applications where it is important to implement threshold functions (such as for neurons), where the thresholding sensitivity is more expressive than a binary component.

## V. CONCLUSION

We have demonstrated a novel in-line spin wave majority gate concept that is both scalable and compatible with conventional CMOS patterning techniques. By individually controlling the phase and amplitude of signal at the three inputs, a full majority truth table was demonstrated. Due to the wave based nature of the operation, the output signal dependent on the applied frequency in an oscillatory way. This could be exploited to demonstrate an inverted majority function in the same device by adjusting the operation frequency. Circuit level benchmarking indicated the high potential of spin wave based devices for low-power electronics.

ACKNOWLEDGMENT

This work was performed as part of the imec IIAP program on Core CMOS and Beyond CMOS. Support from the H2020 project CHIRON (contract No. 801055) is gratefully acknowledged.

REFERENCES

[1] A. Khitun and K. Wang, J. Appl. Phys. 110, 034306 (2011)
[2] Radu et al., Proc. IEEE IEDM (2015)
[3] T. Schneider et al., APL 92, 022505 (2008)
[4] Chumak et al., Nature Comm. Vol. 5 No. 4700 (2014)
[5] T. Fisher et al., APL 110, 152401 (2017)
[6] S. Klinger et al., APL 105, 152410 (2014)
[7] O. Zografos et al., AIP Advances 7 (5), 056020 (2017)
[8] The simulations were performed using "ANSYS HFSS: High Frequency Electromagnetic Field Simulation" software
[9] O. Zografos et al., IEEE NANOARCH 2014, pp. 25-30.
[10] C. Pan et al., IEEE JxCDC 3 2017 101-110.
[11] O. Zografos et al., Chapter 7 - R. Topaloglu and P. Wong, Springer, 2018, ISBN 978-3-319-90384-2.
[12] J. Ryckaert et al., CICC 2014, pp. 1–8.
[13] S. Dutta et al., Scientific reports 7.1 (2017): 17866.

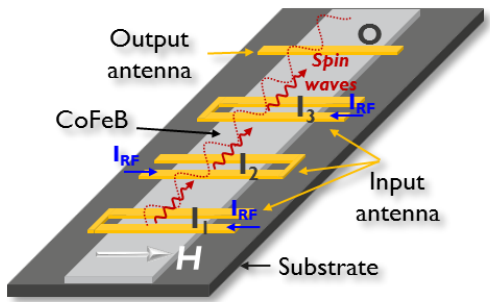

Fig.1. Schematic of a spin wave in-line majority gate

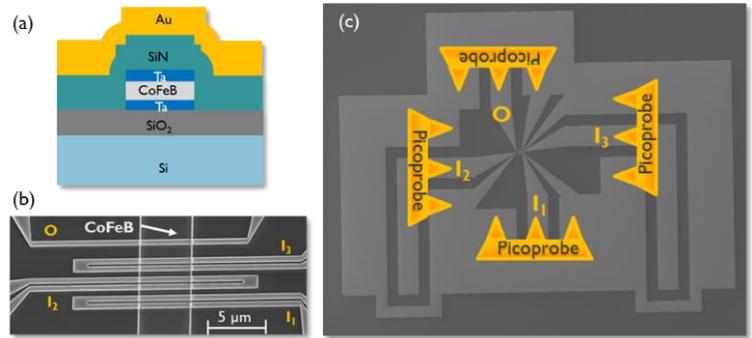

Fig. 2. (a) Schematic of device cross-section under an antenna. (b) SEM image of the active area of the device. The inputs are 3 U-shaped antennas and the output is a single wire antenna. (c) SEM image of the full microwave majority gate device with a sketch of the picoprobe connection to the transducers.

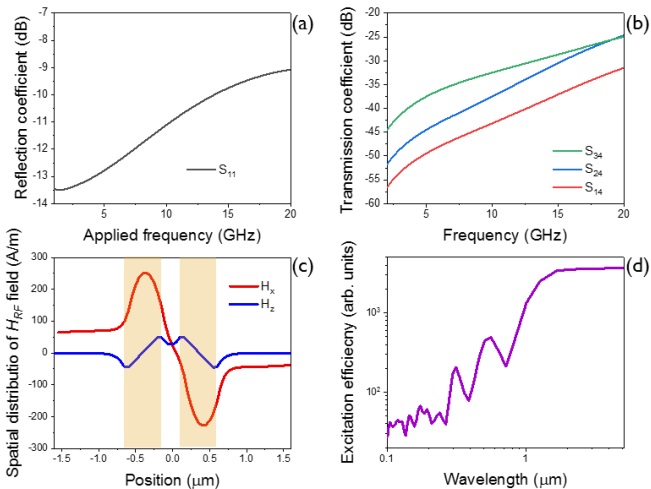

Fig. 3. Scattering parameters of the device simulated by HFSS: (a) reflection coefficient for U-shape antennas, and (b) direct parasitic coupling between every input and the output antenna. (c) Magnetic field components generated by a U-shaped antenna simulated by HFSS and (d) the resulting bandwidth of the device.

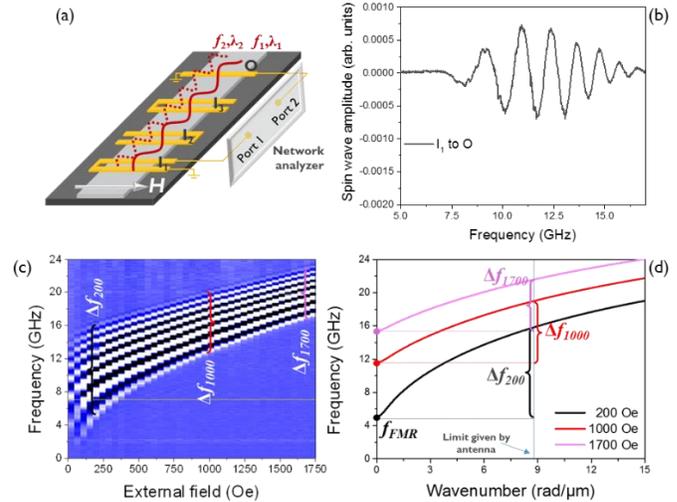

Fig. 4. (a) Schematic of the experiment for a single input and (b) the detected signal due to spin wave propagation at the output. (c) Frequency-field dependence of the spin wave transmission. Light blue color corresponds to zero spin-wave transmission, while the dark blue and the white band stands for propagating spin waves. (d) Spin wave dispersion relation calculated for three values of the magnetic field.

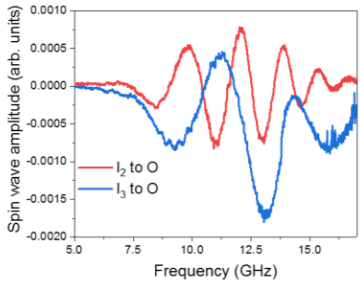

Fig. 5. Spin wave transmission from inputs $I_2$ and $I_3$ to the output O.

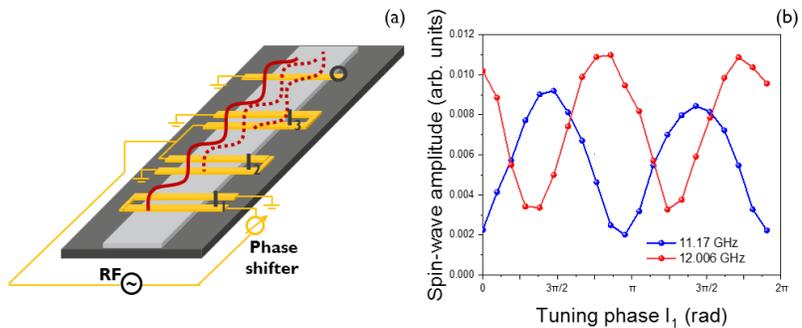

Fig.6. (a) Sketch of a spin wave in-line majority gate showing phase control at input $I_1$. (b) Output signal generated by the interference of the three spin waves. The change of the amplitude is determined by constructive and destructive interference. The bias field was set to 50 mT.

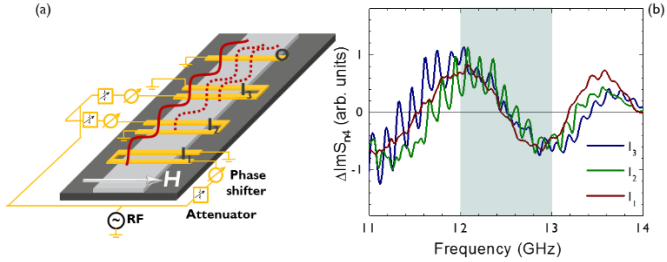
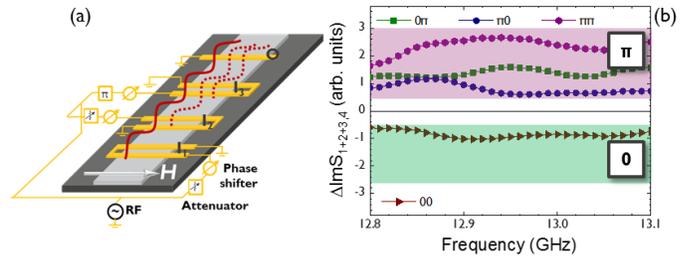

Fig. 7. (a) Schematic of a spin wave in-line majority gate with phase control at each transducer. (b) Output signal generated by each input showing the possibility to match phase and amplitude of the three input spin waves, as required to build logic gates.

Fig. 8. (a) Schematic of an OR gate where one input ($I_3$) is fixed at a phase of $\pi$ as a control gate. (b) Experimental signal demonstrating the functionality as an OR gate.

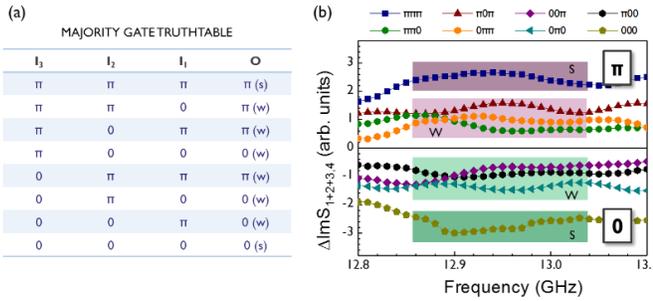
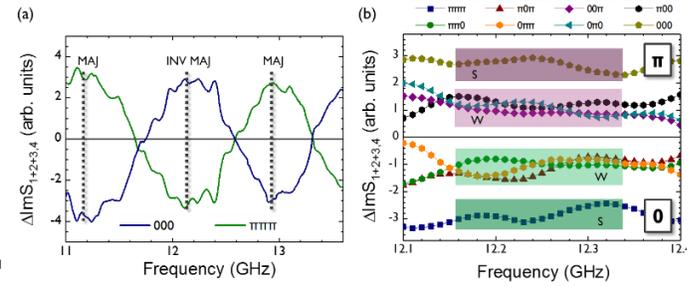

Fig. 9. (a) Majority Gate truth table, indicating cases of strong and weak majority (b) Spin wave transmission due to interference of the 3 waves at the output tranducer. The 8 cases can be observed as well as a clear separation between strong and weak majority gate in a 200 MHz frequency bandwidth.

Fig. 10. (a) Spin wave transmission due to interference of 3 waves at output in a broad frequency span for the two cases of strong majority. Due to the phase rotation it can be observed how a MAJ state transforms in a MIN (INV + MAJ) and then again in MAJ. (b) Experimental MIN truth table where strong and weak minority can be separated in a 200 MHz frequency range.

| NAME | DESCRIPTION |
|---|---|
| CRC32 | Cyclic redundancy check XOR tree |
| BKA264 | 2-operand 64-bit Brent-Kung Adder |
| GFMUL | Mastrovito multiplier for irreducible polynomial |
| CSA464 | 4-operand 64-bit Carry-Skip Adder |
| HCA464 | 4-operand 64-bit Han-Carlson Adder |
| WTM32 | 2-operand 32-bit Wallace tree Multiplier |
| MAC32 | 3-operand 32-bit (7,3) counter tree MAC |
| DTM32 | 2-operand 32-bit Dadda tree Multiplier |
| DIV32 | 2-operand 32-bit Divider |
| DTM64 | 2-operand 64-bit Dadda tree Multiplier |

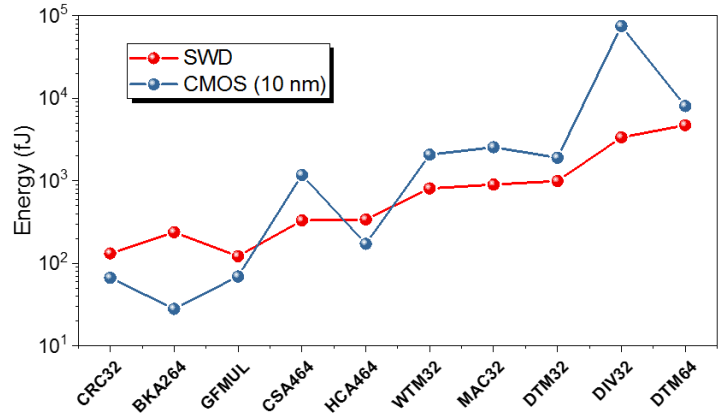

Fig. 11. Descriptions of benchmark designs used to compare spin wave logic and CMOS technologies.

Fig. 12. Energy per operation of spin wave logic circuits and CMOS (10nm) as technology reference. Benchmarks are ordered in increasing circuit size. We observe that for sufficiently large circuits, wave logic operate at lower energy than state-of-the-art CMOS.